\documentclass{elsart4}

\usepackage{graphicx}

\usepackage{amssymb}

\usepackage[english,francais]{babel}

\newtheorem{e-proposition}[theorem]{Proposition}

\newtheorem{e-definition}[theorem]{Definition\rm}

\renewcommand{\theequation}{\arabic{equation}}
\def\og{\leavevmode\raise.3ex\hbox{$\scriptscriptstyle\langle\!\langle$~}}
\def\fg{\leavevmode\raise.3ex\hbox{~$\!\scriptscriptstyle\,\rangle\!\rangle$}}

\newcommand\epeak{E$_\mathrm{peak}$ }
\newcommand\eiso{E$_\mathrm{iso}$ }

\begin{document}

\centerline{The prompt emission of GRBs}
\begin{frontmatter}


\selectlanguage{english}
\title{Observing the prompt emission of GRBs}


\selectlanguage{english}
\author[jla]{Jean-Luc Atteia},
\ead{atteia@irap.omp.eu}
\author[mb]{Michel Bo\"er}
\ead{Michel.Boer@oamp.fr}

\address[jla]{IRAP, Universit\'e de Toulouse (UPS-OMP), CNRS, 14 Avenue E. Belin, 31400 Toulouse, France}
\address[mb]{Observatoire de Haute-Provence (CNRS), F 04870 Saint Michel l'Observatoire, France}


\medskip

\begin{abstract}
Gamma-ray bursts (GRBs) were first detected thanks to their prompt emission, which was the only information available for decades. In 2010, while the high-energy prompt emission remains the main tool for the detection and the first localization of GRB sources, our understanding of this crucial phase of GRBs has made great progress. We discuss some recent advances in this field, like the occasional detection of the prompt emission at all wavelengths, from optical to GeV; the existence of sub-luminous GRBs; the attempts to standardize GRBs; and the possible detection of polarization in two very bright GRBs. Despite these advances, tantalizing observational and theoretical challenges still exist, concerning the detection of the faintest GRBs, the panchromatic observation of GRBs from their very beginning, the origin of the prompt emission, or the understanding of the physics at work during this phase. 

Significant progress on this last topic is expected with SVOM thanks to the observation of dozens of GRBs from optical to MeV during the burst itself, and the measure of the redshift for the majority of them. SVOM will also change our view of the prompt GRB phase in another way. Within a few years, the sensitivity of sky surveys at optical and radio frequencies, and outside the electromagnetic domain in gravitational waves or neutrinos, will allow them to detect several new types of transient signals, and SVOM will be uniquely suited to identify which of these transients are associated with GRBs. This radically novel look at GRBs may elucidate the complex physics producing these bright flashes.


\vskip 0.5\baselineskip

\selectlanguage{francais}
\noindent{\bf R\'esum\'e}
\vskip 0.5\baselineskip
\noindent
{\bf Observer l'\'emission prompte des sursauts gamma. }
Les sursauts gamma ont \'et\'e initialement d\'etect\'es gr\^ace \`a leur \'emission prompte, qui a constitu\'e pendant quelques d\'ecennies la seule information disponible sur ces \'ev\'enements. En 2010, tandis que le ``flash gamma'' reste le seul moyen de d\'etecter les sursauts gamma cosmiques et de fournir une premi\`ere position sur le ciel, notre compr\'ehension de cette phase cruciale des sursauts a fait des progr\`es importants. Nous pr\'esentons quelques avanc\'ees r\'ecentes dans ce domaine, comme la d\'etection occasionnelle de l'\'emission prompte \`a toutes les longueurs d'onde, du visible au GeV ; l'existence d'une classe de sursauts sous-lumineux ; les tentatives de standardisation des sursauts gamma ; et la d\'etection possible de la polarisation dans deux sursauts tr\`es brillants. Malgr\'e ces avanc\'ees, les d\'efis observationnels et th\'eoriques restent nombreux pour comprendre cette phase, ils concernent notamment la d\'etection des sursauts les plus faibles, l'observation panchromatique des sursauts depuis le d\'ebut de l'\'emission, l'origine de l'\'emission prompte et la compr\'ehension des processus physiques \`a l'\oe uvre pendant cette phase.

Sur ce dernier point, des progr\`es importants sont attendus avec SVOM, gr\^ace \`a l'observation de plusieurs dizaines de sursauts du visible au MeV pendant la phase prompte, et \`a la mesure du redshift de la majorit\'e d'entre eux. SVOM est en outre appel\'e \`a changer notre vision des sursauts gamma d'une autre mani\`ere. D'ici quelques ann\'ees la sensibilit\'e des grands programmes de surveillance du ciel en visible et radio, mais aussi en dehors du domaine \'electromagn\'etique, dans les ondes gravitationnelles ou les neutrinos, sera suffisante pour permettre la d\'etection de nouveaux types de signaux transitoires, et SVOM sera id\'ealement plac\'e pour identifier ceux qui sont associ\'es \`a des sursauts gamma. Cette perception nouvelle de ces \'ev\'enements apportera un \'eclairage in\'edit sur la physique complexe qui produit ces flashs intenses de rayons gamma.
%

{\it Pour citer cet article~: J-L. Atteia, M. Bo\"er, C. R. Physique ? (2011).}
\vskip 0.5\baselineskip
\noindent{\small{\it Mots-cl\'es~:} Astrophysique des hautes Žnergies ; Instrumentation spatiale : SVOM ; Sursauts gamma ; Trous noirs ; Astrophysique multi messagers}}

\end{abstract}
\end{frontmatter}


\selectlanguage{english}
\section{Introduction: An historical perspective on prompt emission}
\label{s-intro}

Observationally, the prompt GRB has been defined as the bright flash of high-energy photons (keV -- MeV) triggering GRB detectors in space. This emission, lasting between 10 milliseconds and 1000 seconds (see figure \ref{fig-t90}), has played a central role in the study of GRBs since the announcement of their discovery by Klebesadel, Strong \& Olson in 1973 \cite{Klebesadel1973}. During nearly 25 years, until the discovery of the afterglows in 1997 \cite{Costa1997}, the prompt emission was the only information available to astronomers. It provided the first clues on the origin of GRBs, revealing their isotropy on the sky, the lack of bright persistent counterparts and their non-thermal spectrum. It also pointed out the existence of two classes of GRBs: the short and the long ones, with a separation around two seconds \cite{Kouveliotou1993}, both classes being distributed isotropically on the sky.

\begin{figure}[htbp]
\begin{center}
\includegraphics[width=8cm]{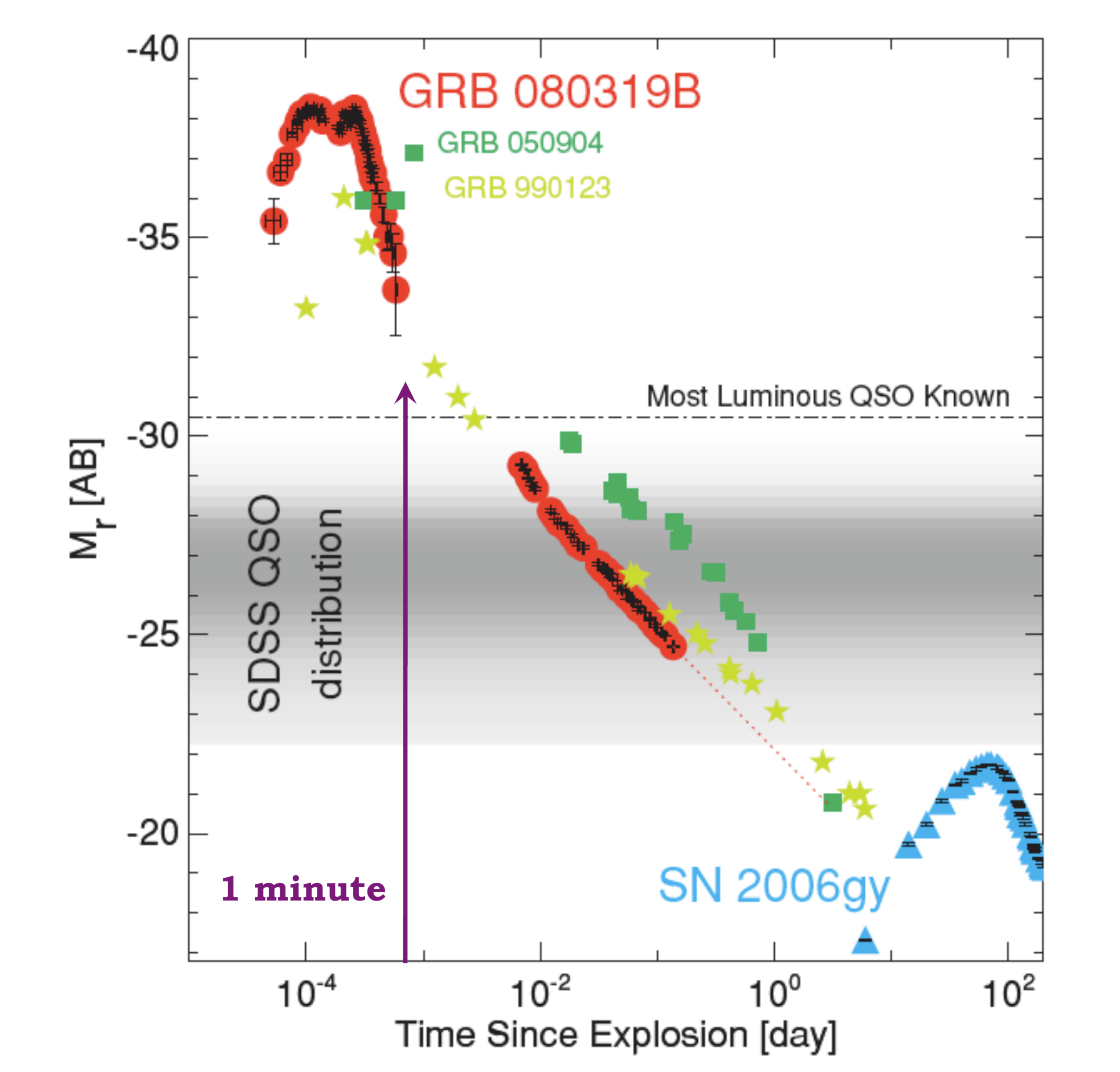}
\caption{Evolution of the luminosity of GRB~080319B, GRB~050904 and GRB~990123 at visible wavelengths. The transition between the prompt phase and the early afterglow of GRB~080319B is clearly seen at time $\sim$1 min. For comparison, the figure shows the luminosities of quasars (QSOs) and of the very bright supernova SN 2006gy -- from Bloom et al. (2009).}
\label{fig-grb080319b}
\end{center}
\end{figure}

The puzzle that prevented the early realization of the cosmological distances of GRBs was the tremendous amount of energy released in the few seconds of the prompt GRB when extragalactic distances were considered. This fantastic energy release, combined with the sub-second luminosity fluctuations of GRBs, implied a huge photon density within a very small volume. This would lead to high opacity for photon-photon absorption and the complete absorption of photons above 0.5 MeV, in contrast with the observations \cite{Schmidt1978}. This paradox was solved with the development of models producing GRBs within relativistic jets pointing in our direction. When Lorentz factors of several tens to several hundreds are considered, the radiation from the jet is greatly amplified by the Doppler boost. In such ``fireball models" the prompt emission is produced within the jet, while the afterglow is due to the interaction of the jet with the matter surrounding the source. The visible emission of GRB~080319B (fig. \ref{fig-grb080319b}) illustrates these two phases with a bright flare lasting  $\sim$1 min, associated with the prompt emission, followed by a fainter and smoothly decaying afterglow. The mechanisms leading to energy dissipation within the jet and the dominant radiation processes are still debated: relativistic shocks and/or magnetic reconnection are frequently advocated to dissipate the internal energy of the jet, which is then radiated as synchrotron and/or blackbody radiation. The detection of the prompt gamma-rays implies that they are emitted after the jet has become transparent to its own radiation. When the jet becomes transparent, most of its thermal energy should be radiated away as photospheric emission. This thermal emission is not commonly observed, implying that the energy is temporarily stored in non-thermal components: kinetic energy, magnetic fields... and released after the jet has become transparent.

Despite the general agreement about this picture, many issues stand unsolved. Theoretical questions remain about the origin of the prompt emission, the importance of photospheric radiation, the role of pairs/baryons/magnetic fields, the microphysics, etc. On the observational side, the diversity of the prompt emission is such that we suspect that current instruments could miss a significant fraction of the bursts, the multi-wavelength spectrum of the prompt emission is extremely difficult to measure, the relativistic motion of the jet and the panchromatic nature of the emission complicates the identification of the origin of the observed components...
The broad range of GRB distances contributes to confuse the picture: \textit{i)} it biases the detection of intrinsically faint sources; \textit{ii)} cosmological effects (time dilation and spectral softening) are strong for distant bursts, making GRBs practically useless when their distance is unknown, and complicating statistical studies; \textit{iii)} the role of cosmological evolution and its effect on GRB light-curves and spectra remains to be assessed.
Finally, the impact of the physical parameters of the source on the properties of the prompt emission (duration, brightness, peak energy...) is not elucidated.

In the following, we discuss the role of the prompt emission for the detection of GRBs, the progress achieved in the Swift era, and the expectations of SVOM. At that point, we should emphasize the contribution of instruments currently observing prompt emission on a wider spectral range than Swift/BAT \cite{Barthelmy2005}, like Konus \cite{Aptekar1995}, Suzaku WAM \cite{Yamaoka2009}, or Fermi GBM \cite{vonKienlin2004}. These instruments permit measuring  the distribution and evolution of GRB spectral parameters (e.g. \cite{Krimm2009}, \cite{Sakamoto2010}), they allow correlation studies (see section \ref{ss-standard} below), and they facilitate the distinction between long and short GRBs (\cite{Ghirlanda2009}, \cite{Goldstein2010}, \cite{Lu2010}, \cite{Nava2010}).

\section{Using the prompt emission for GRB detection and localization}
\label{s-detection}

\begin{figure}[htbp]
\begin{center}
\includegraphics[width=8cm]{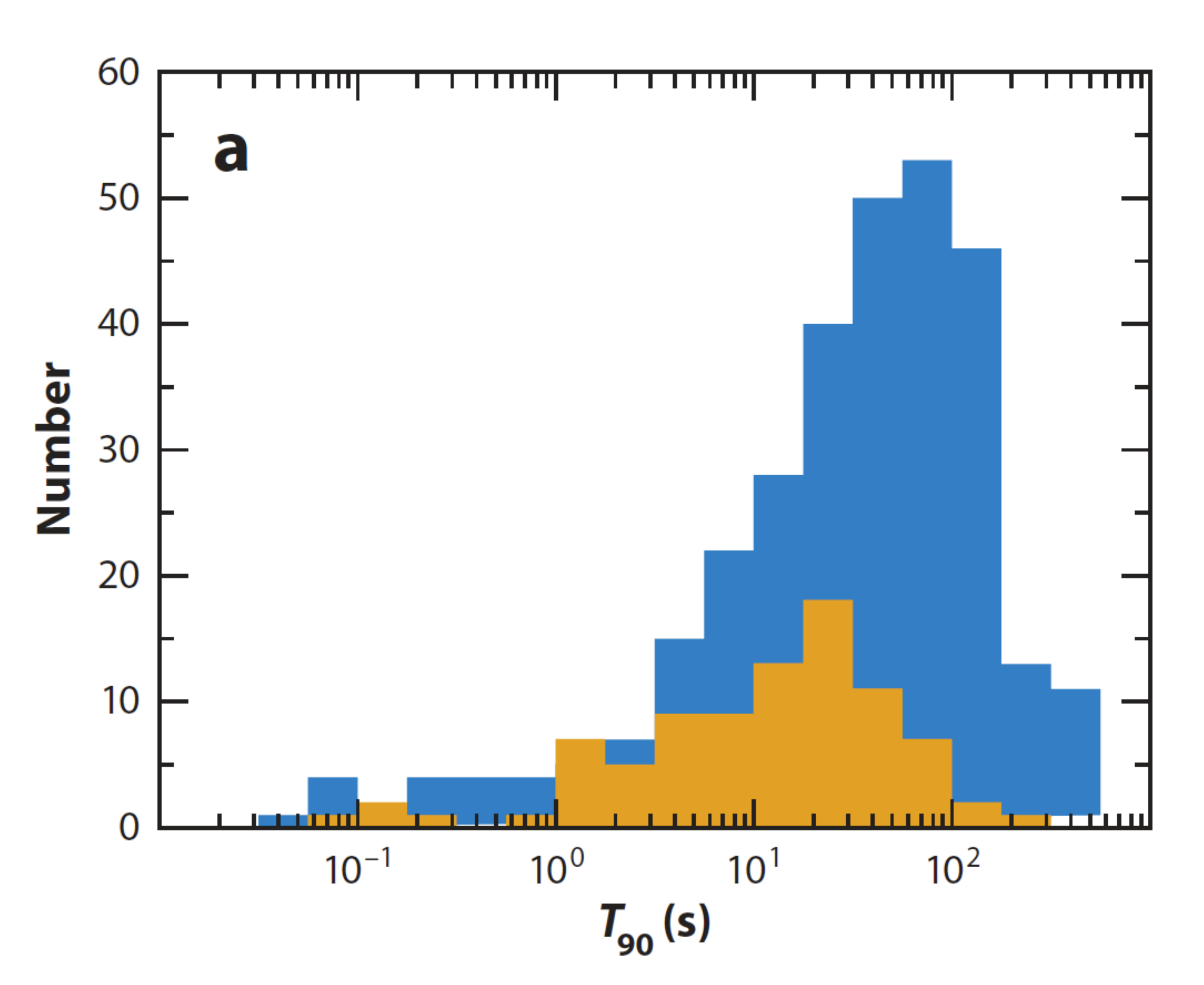}
\caption{Histogram of the duration of the prompt emission of Swift GRBs -- from Gehrels et al. (2009); the blue distribution is the observed $\rm T_{90}$ distribution, while the orange one is corrected for source redshift.}
\label{fig-t90}
\end{center}
\end{figure}

Historically, it is the detection of the prompt emission in hard X-rays / gamma-rays which led to the discovery of GRBs. Interestingly, 40 years later this method remains the one which has led to the discovery of all known GRBs. While the situation will probably change with the advent of deep and wide optical surveys (e.g. the Large Synoptic Survey Telescope), or wide field IR or X-ray detectors in space, well suited for the detection of the afterglows in the first day following a burst.

\begin{figure}[htbp]
\begin{center}
\includegraphics[width=10cm]{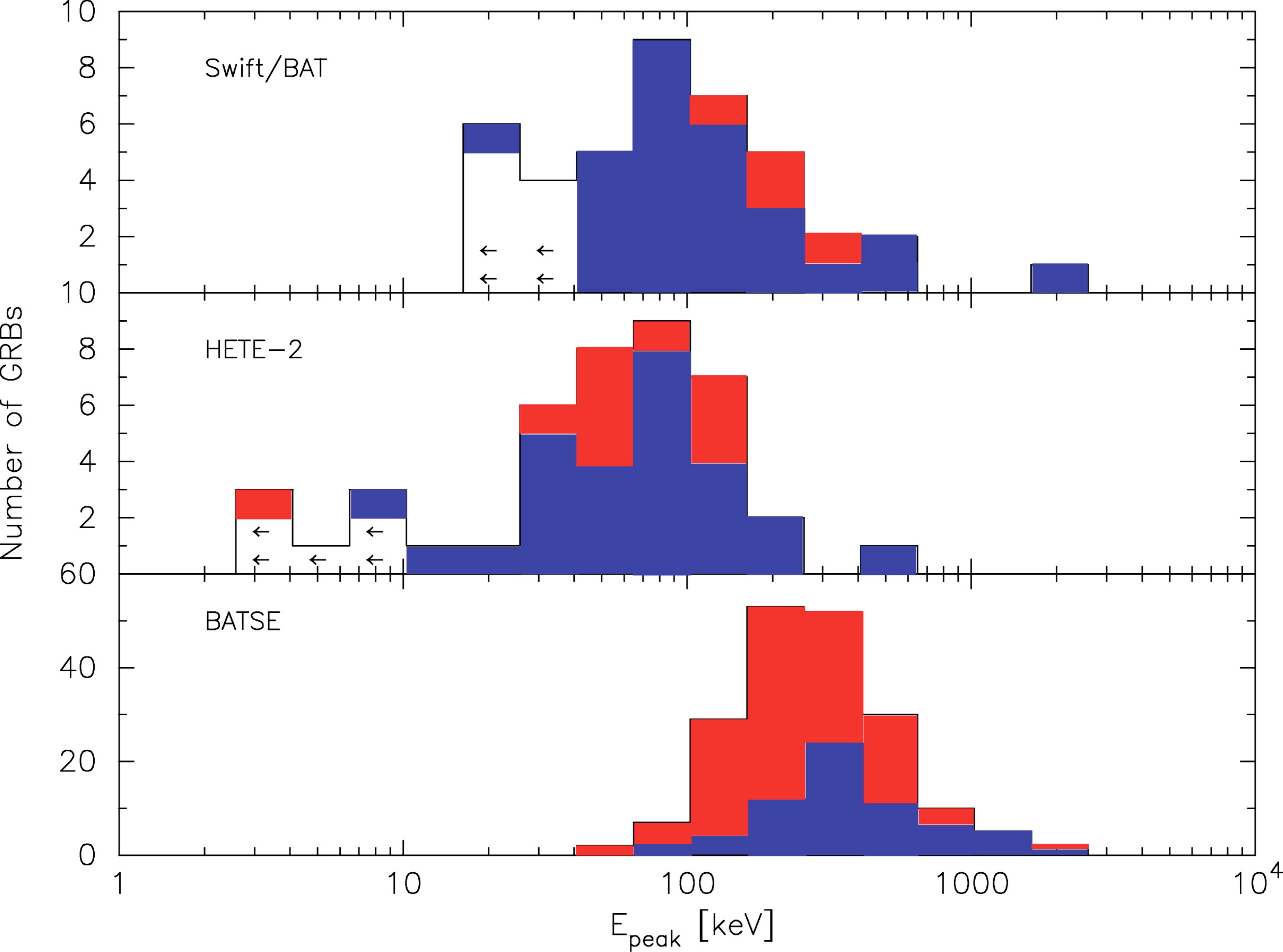}
\caption{The distribution of \epeak measured with Swift/BAT (top), HETE-2 (middle), and BATSE (bottom), from Sakamoto et al. (2008). The low energy bound of the distribution is correlated with the low energy threshold of the trigger -- 15 keV for Swift, 4 keV for HETE-2, and 50 keV for BATSE (see http:\textbackslash\textbackslash www.batse.msfc.nasa.gov\textbackslash batse\textbackslash grb\textbackslash catalog\textbackslash current\textbackslash tables\textbackslash batse.trig\_crit) -- suggesting the existence of a population of bursts with low \epeak which is incompletely sampled by current instruments. For comparison, the energy ranges of the instruments are 15-250 keV for Swift, 4-400 keV for HETE-2, and 25-1000 keV for BATSE.}
\label{fig-epeak}
\end{center}
\end{figure}

Over the years, different instruments have been used to detect GRBs. These instruments have emphasized the role of various factors which shape, and bias, the populations of detected GRBs:
\begin{itemize}
\item The evolution of the instruments from simple spectro-photometers to more complex spectro-imagers led to a decrease of the fraction of short GRBs, from $\sim$25\% with BATSE to below 10\% with Swift (fig. \ref{fig-t90}). This is explained by the need of having a minimum number of photons to make an image with a coded mask and by the fact that this minimum number of photons is more difficult to get for short GRBs.
\item The instruments with a low energy threshold at few keV have discovered ``X-ray only" GRBs, with few photons above 30 keV (\cite{Heise2001} ; \cite{Barraud2003} ; \cite{Sakamoto2005} ; \cite{Sakamoto2008}). These events, which have been called X-Ray Flashes, show that we have not yet reached the {\bf lower} limit of the \epeak distribution 
    (see Fig. \ref{fig-epeak}).
\item The increasing size of the detectors led to the discovery of few ``sub-luminous GRBs", which are typically 10$^{2-4}$ times fainter and 10$^{1-2}$ times closer than classical GRBs. These events, which could constitute the bulk of the GRB population, are discussed in more details in section \ref{ss-subluminous} below.
\end{itemize}

Considering GRB detection, the combination of these instrumental effects with the evolution of GRB properties with distance (due to cosmological effects and intrinsic evolution of the sources) gives us a biased view of the GRB populations present in the universe and of their relative importance.

In term of localization capability, GRB missions have evolved quickly after BATSE and BeppoSAX. All GRB missions (HETE-2, Swift, as well as future missions) now have the capability to provide rapid ($<$60 sec) arcminute localizations. This is also true for several general purpose gamma-ray missions like INTEGRAL, Agile, and Fermi/LAT. The localization of the GRB is usually computed on-board, with the fast deconvolution of a coded mask image containing a single source: the GRB. The localization can be sent to the ground via a relay satellite, if the GRB satellite is equipped to communicate with it (Swift), or via direct communication with the ground (HETE-2, INTEGRAL, SVOM). Satellites on a low earth orbit are only visible for few minutes from a given point of the earth, requiring a network of alert receiving stations (14 for HETE-2 with $i \sim 0^\circ$, and nearly 40 for SVOM with $i \sim 30^\circ$). 

\section{The prompt emission in the Swift era}
\label{s-swift}

\begin{figure}[htbp]
\begin{center}
\includegraphics[width=8cm]{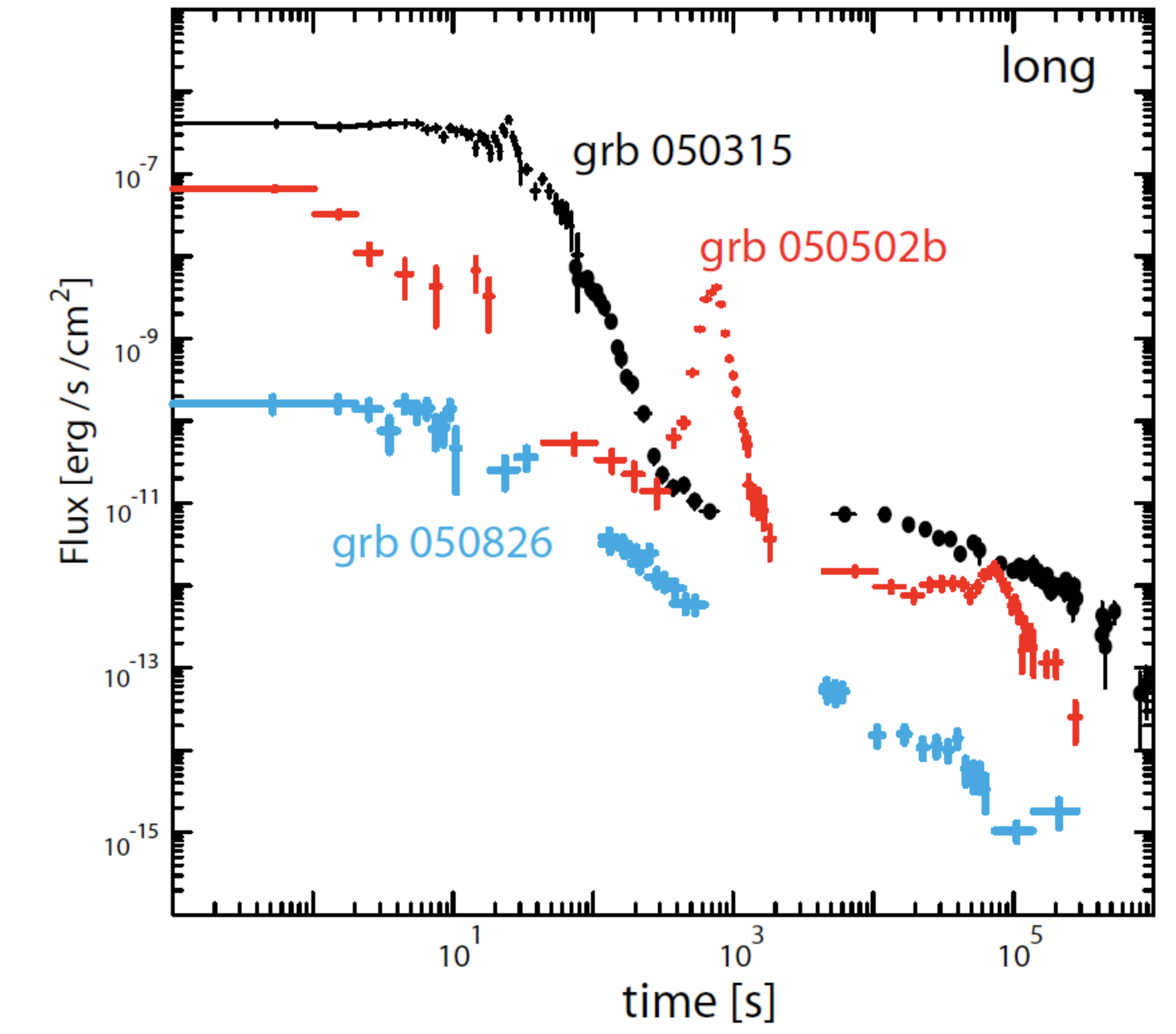}
\caption{Three GRB light-curves in the hard X-ray range obtained by combining data from Swift/BAT (15-150 keV) and Swift/XRT (0.2-10 keV). GRB~050315 shows a clear separation between the prompt emission, which ends with an exponential decay around T $\sim$ 100 sec, and a plateau phase preceding the decaying afterglow. GRB~050826 shows a smooth transition between the prompt emission and the afterglow at T $\sim$ 10 sec. GRB~050502B shows a very bright X-ray flare at T $\sim$ 700 sec superimposed on the canonical decay of the afterglow -- From Gehrels et al. (2009).}
\label{fig-swiftLCs}
\end{center}
\end{figure}

Even if the prompt emission is the component of GRBs which has been studied for the longest time, recent observations have significantly changed our view of this crucial phase of GRBs. We briefly discuss here these observations and current issues concerning the observation of the prompt phase of GRBs. We do not address theoretical issues which are discussed in this volume by B. Zhang and Lemoine \& Pelletier, nor the High Energy emission, which is treated by Piron \& Connaughton.

\subsection{Defining the prompt emission...}
\label{ss-duration}
The first question is the definition of the prompt emission. We mentioned that it was initially defined observationally as the bright flash of high-energy photons easily detected by space detectors.  Given the distance of GRBs, opacity constraints require this radiation to be emitted within a jet moving toward us with relativistic velocity, thus providing a theoretical framework for the definition of the prompt emission. It is also commonly admitted that the slowly decaying afterglow is due to the shock of the jet on the circumstellar/interstellar medium. This simple view faces new challenges with the observations of Swift. 

One challenge is the complexity of the X/gamma-ray light-curves (Fig. \ref{fig-swiftLCs}), which rarely show a simple transition between the bright prompt emission and the fainter afterglow, as one would have expected for such different origins. The exploration of the transition period between the prompt emission and the afterglow, and the discovery of the spectro-temporal complexity of this phase is one of the major results of Swift. For completeness, it should be noted that this transition was studied before Swift in some bright GRBs (see for instance \cite{Costa1998}, \cite{Burenin1999}, \cite{Giblin1999} ; \cite{Connaughton2002} for long bursts and \cite{Lazzati2001} for short bursts) and that hints of the complexity of the prompt emission were already pointed out, like the existence of quiescent periods lasting $>$100~sec in some GRBs (\cite{RamirezRuiz2001a}, \cite{McBreen2006}, \cite{Romano2006}), or the presence of late X-ray flares in two bright GRBs detected with BeppoSAX \cite{Piro2005}. 

We are thus facing a difficulty to define the prompt emission. For observers, the prompt phase is defined by the detection of the burst with high-energy detectors (though the prompt emission is seen at all wavelengths), and as such it depends strongly on the sensitivity and energy range of these detectors. Alternatively, one could associate the prompt emission with the radiation emitted within the jet (the proper activity of the jet -- internal shocks or magnetic reconnection -- and the reverse shock). In this case the prompt emission is probably much longer and superimposed on the afterglow since the X-ray flares discovered by Swift in many GRBs in the minutes following the burst (see \cite{Tagliaferri2005} and GRB~050502B in fig. \ref{fig-swiftLCs}) have been attributed to activity taking place within the jet (long lasting activity of the central engine or shells emitted with smaller Lorentz factors).

After 1999 it has been possible to measure the optical emission during the prompt phase of few GRBs, however the optical light-curves do not show a clear pattern: in some bursts they appear correlated with the high-energy emission, and in other bursts they don't. In most GRBs for which we have early optical observations, there is no clear transition between the prompt phase and the afterglow (except for GRB~080319B, as explained in section \ref{ss-optical} below).

Considering short GRBs, the confusion is also present. The discovery of $\sim$100 sec long soft emission following the initial short pulse of GRB~050709 \cite{Villasenor2005}, and the detection with Swift of two long nearby GRBs without a supernova (a characteristic of short GRBs) \cite{DellaValle2006} \cite{GalYam2006}, raised the question of the identification of short GRBs on the sole basis of their prompt emission \cite{Zhang2007} or duration.

\subsection{GRB standardization}
\label{ss-standard}

\begin{figure}[htbp]
\begin{center}
\includegraphics[width=10cm]{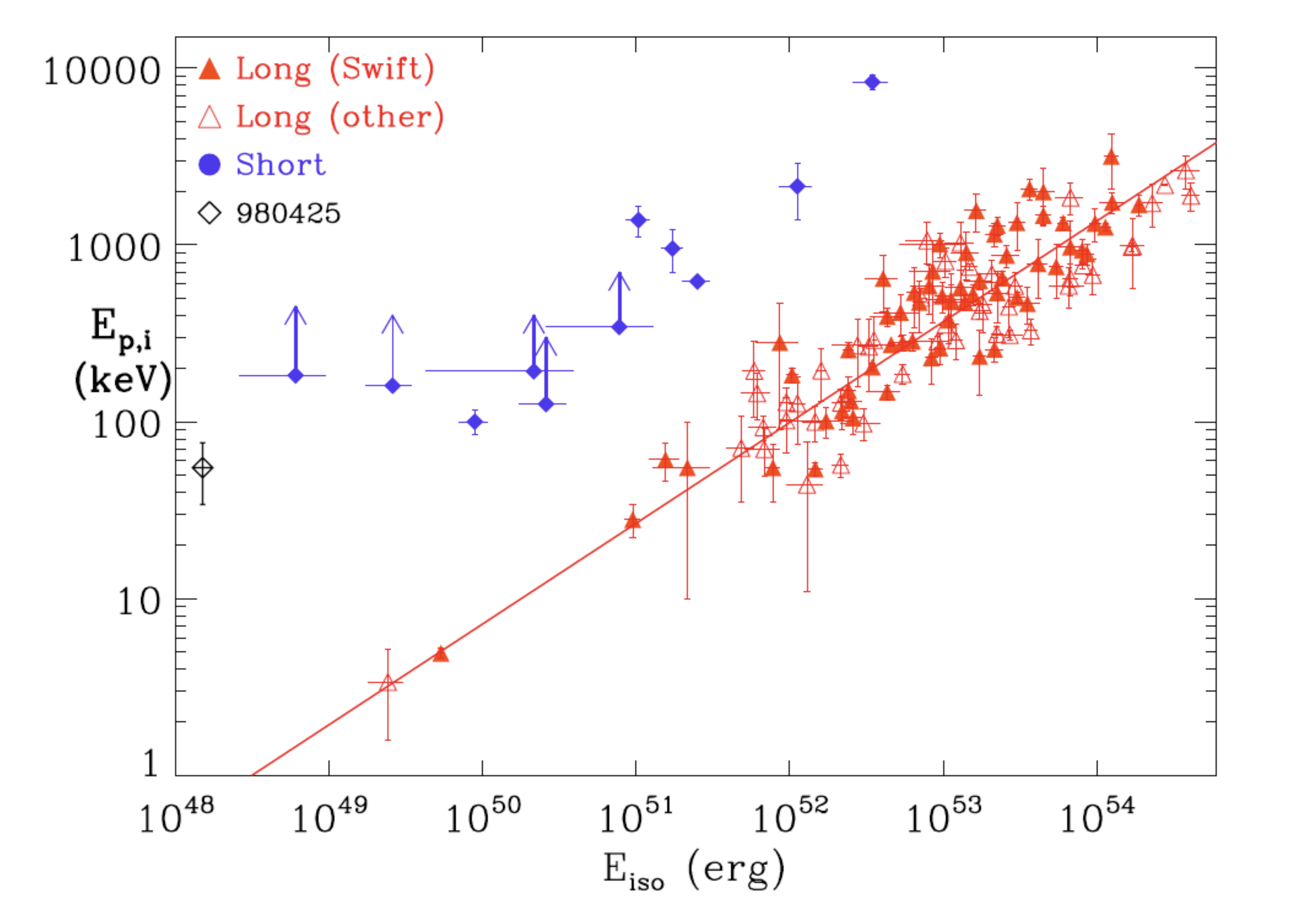}
\caption{The intrinsic peak energy (the maximum of the SED) as a function of the isotropic energy (the energy radiated by the GRB assuming isotropic emission) for a large sample of long and short GRBs whose redshift has been measured \cite{Amati2010}. This figure shows that bright GRBs (large \eiso) have more energetic photons (large E$_{\rm p,i}$) on average -- From Amati (2010).}
\label{fig-amati}
\end{center}
\end{figure}

In the early times, when only the prompt emission of GRBs was measured, it was noticed by GRB observers that the hardness of GRBs (the fraction of high-energy photons) was correlated with their intensity (\cite{Mallozzi1995}, \cite{Mitrofanov1996}, \cite{Dezalay1997}). With the measure of the first redshifts, it became possible to compute the intrinsic properties of GRBs, especially their true energetics (luminosity ; total energy, \eiso ...). Amati et al. (2002) found a strong correlation between the peak of the spectral energy distribution (E$_{\rm p,i}$) and the energy output assuming isotropic emission (\eiso) of 11 GRBs with a redshift detected with BeppoSAX. This correlation, which was later confirmed with the detection of many more GRBs (see Fig. \ref{fig-amati} \cite{Amati2010}), can be used to 'standardize' GRBs: to infer the true energy output of a burst (or its luminosity) when its redshift and E$_{\rm p,i}$ are known. Other relations allow the standardization of GRBs: limiting the discussion to the prompt emission, the correlation of spectral lags \cite{Norris2000}, or variability \cite{Reichart2001} with the intrinsic luminosity was previously known, but the scatter of these relations is larger than the scatter of the Amati relation. 

Such relations allow the construction of a {\it Hubble diagram}\footnote{the Hubble diagram shows the luminosity distance as a function of redshift for cosmological sources.} of GRBs, which depends directly on the cosmological parameters. In the case of Type Ia supernovae, the construction of the Hubble diagram enabled the discovery of the acceleration of the universe at the end of the 90's \cite{Riess1998}, \cite{Perlmutter1999}.
However, the dispersion of the relations used for GRB standardization is large and their calibration is still debated. These limitations prevent using individual GRBs to construct the Hubble diagram, which can only be constructed on a statistical basis\footnote{This is the same lack of accuracy which prevents using GRB standardization to compute the redshift of individual GRBs when optical spectroscopy is not possible.}, this explains why the GRB Hubble diagram is significantly less accurate than the SN one. 
The afterglow at X-rays, optical or infrared wavelengths can also be used to standardize GRBs, though this does not result in better constraints for the moment.
An interesting attempt to reduce the uncertainty of the GRB Hubble diagram has been done by Schaefer (2007). Instead of searching the 'best' luminosity indicator, he considers several  low quality luminosity indicators and computes the GRB luminosity as the average of all the different luminosities given by all the indicators \cite{Schaefer2007}. 

Despite these limitations many authors have studied the constraints that current (and future) GRB samples (could) impose on cosmological parameters, considering that GRBs are complementary to type Ia supernovae because they can be detected up to very high redshifts (z$\sim$8 at least), and because the selection effects are different. This field of studies is very active and SVOM may be able to bring significant progress with the good characterization of GRBs that it will permit and the expected large fraction of GRBs with a redshift. A good introduction to the use of GRBs as standard candles has been given by Ghirlanda et al. (2006) \cite{Ghirlanda2006}.

\subsection{Sub-luminous GRBs}
\label{ss-subluminous}

\begin{figure}[htbp]
\begin{center}
\includegraphics[width=8cm]{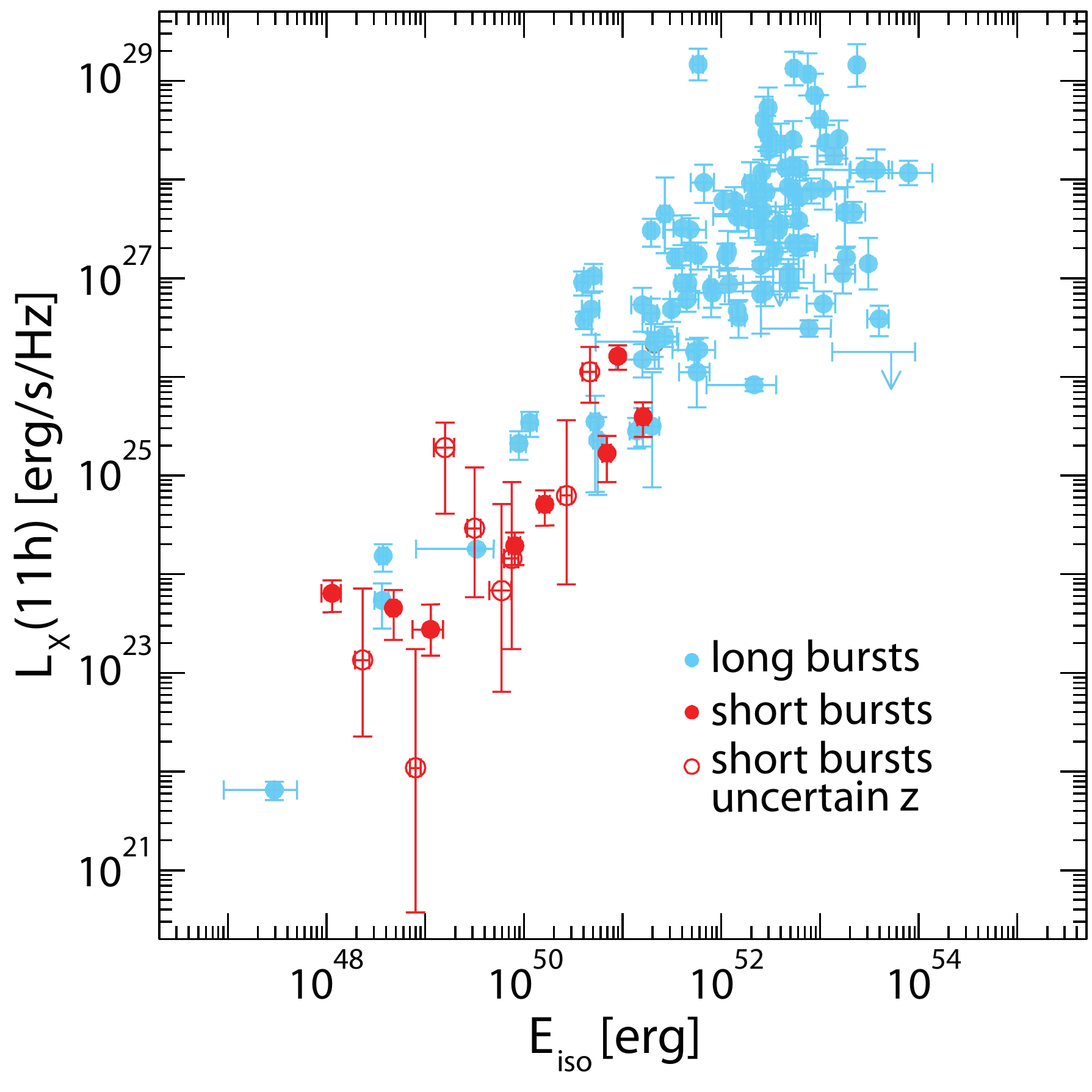}
\caption{Luminosity of the X-ray afterglow as a function of the energy radiated during the prompt phase for Swift GRBs with a redshift. This figure illustrates the very broad range of luminosities of Swift GRBs and the existence of GRBs 10$^{4-5}$ fainter than average which can only be detected in a volume much smaller than classical GRBs (see text) -- From Gehrels et al. (2009).}
\label{fig-swift-eiso}
\end{center}
\end{figure}

Figure \ref{fig-swift-eiso} displays \eiso and the luminosity of the X-ray afterglow at 11 hrs for Swift GRBs whose redshifts have been measured \cite{Gehrels2009}, showing that both quantities span a very broad range (even if evidences have been found for a common behavior of afterglow X-ray light curves \cite{BG2000}\cite{GB2005}, as well as optical \cite{LZ2006} and infrared \cite{GB2008}). Since GRBs are detected from their prompt emission, the broad luminosity function of E$_\mathrm{iso}$, extending from below 10$^{48}$ to nearly 10$^{54}$ erg, will have a strong impact on the volume into which GRBs can be detected: faint GRBs will only be detected locally (z~$\le 0.05-0.1$), while bright GRBs will be detected across the entire universe\footnote{In reality the detection of GRBs depends on the number of photons in the energy range of the detector and not of the total energy. For the purpose of this discussion, however, it is safe considering \eiso as a good proxy of the number of photons and as an indicator of the volume into which a GRB can be detected.}, though cosmological effects may temper this statement. The observed population is thus strongly biased in favor of the bright end of the luminosity function, while GRBs releasing less than $10^{50}$ erg are greatly under-sampled. 

The sample of sub-luminous bursts is made of events which have all been associated with supernovae of type Ib or Ic: GRB~980425 looks like a classical GRB detected by BATSE and BeppoSAX \cite{Soffitta1998}, GRB~020903 is a very soft burst detected with HETE-2 \cite{Sakamoto2004}, GRB~031203 is a classical GRB detected with INTEGRAL \cite{Sazonov2004}\cite{Watson2004}, GRB~060218 and GRB~100316D are long soft bursts detected with Swift \cite{Campana2006}\cite{Starling2010}. The main properties of the prompt emission of these bursts are summarized in Table \ref{tab-ss-lumi}. After the detection of GRB~060218, Soderberg et al. (2006) computed the burst rate of sub-luminous GRBs to be 100--300 Gpc$^{-1}$yr$^{-1}$, which has to be compared with a rate of $\sim$0.5 Gpc$^{-1}$yr$^{-1}$ for classical GRBs \cite{Schmidt2001}\cite{Guetta2005}. These rates suggest that low-luminosity bursts could constitute the bulk of the GRB population, which is barely scratched by current instruments. This possibility shows the importance of developing sensitive instruments for the detection of long, soft bursts in the future.

\begin{table}[htp]
\centering
\begin{tabular}{|l|c|c|c|c|}
\hline
GRB Name & redshift & Duration (T$_{90}$) & E$_{\rm peak, obs}$ & \eiso \\
&  & second & keV & erg \\ 
\hline 
GRB\ 980425      & 0.0085    &  30      &  55               & $1.10^{48}$   \\ 
GRB\ 020903      & 0.25        & 10       &  3                 & $3.10^{49}$   \\ 
GRB\ 031203      & 0.106      & 3         &  ~$<$20 or $\sim$200$^\ast$~  &  $5.10^{49}$  \\ 
GRB\ 060218      & 0.0335    & 2100   &  5                 &  $6.10^{49}$  \\ 
GRB\ 100316D~ & 0.0591    & $>$1300  &  $\sim$30    &  ~$6.10^{49}$~  \\ 
\hline
\end{tabular}

$^\ast$ The E$_{\rm peak, obs}$ of this burst is controversial (see \cite{Sazonov2004} and \cite{Watson2004}).
\caption{Main properties of the prompt emission of 5 sub-luminous GRBs}
\label{tab-ss-lumi}
\end{table}

\subsection{The polarization of the prompt emission}
\label{ss-polar}
Polarimetry offers a powerful way of assessing the nature of the GRB radiation mechanisms, the geometry of the jet, and the origin of the prompt emission. Currently, there is no high-energy mission in operation, which is specifically designed to measure the polarization of the prompt GRB emission at energies of tens to hundreds of keV, nevertheless attempts have been made with non-GRB missions having some capabilities for measuring the polarization of gamma-rays.

Coburn \& Boggs (2003) have reported the detection of linear polarization from the bright GRB~021206 with a polarization level of 80$\pm$20\% \cite{Coburn2003}. This claim has been however refuted by Wigger et al. (2004) who find a level of polarization of 41$^{+57}_{-44}$\% \cite{Wigger2004} and conclude that the quality of the data is insufficient to constrain the level of polarization, and by Rutledge \& Fox (2004) who find that the number of scattered photons (those which are useful for the measure of the polarization) is too low to constrain the polarization \cite{Rutledge2004}.

McGlynn et al. (2007) have studied the linear polarization of GRB~041219A, the brightest GRB detected with INTEGRAL, with the spectrometer SPI \cite{McGlynn2007}. They find weak evidence for polarization during the brightest pulse of duration 66 s, with a level of linear polarization of 63$\pm$30\% at an angle of 70$\pm$14 degrees in the 100-350 keV energy range. They also constrain degree of polarization in the brightest 12 s of the GRB, finding a polarization fraction of 96$\pm$40\% at an angle of 60$\pm$14 degrees over the same energy range. However, despite extensive analysis and simulations, they could not definitively exclude a systematic effect that could mimic the weak polarization signal they measure. This is also the conclusion of Kalemci et al. (2007), who find a polarization fraction averaged over the entire burst, of 98$\pm$33\% at an angle of 45$\pm$5 degrees over the energy range 100-350 keV, but indicate that they cannot strongly rule out the possibility that the measured modulation is dominated by instrumental systematics \cite{Kalemci2007}. Finally, G\"otz et al. (2009) have analyzed the polarization of GRB~041219A, with IBIS the imager of INTEGRAL, in the energy range 200-800 keV \cite{Gotz2009}, reaching mostly the same conclusion. They find no polarization in the first (brightest) peak and an average fraction of 43$\pm$25\% for the second peak. Time resolved analysis within the peaks shows large, rapid variations of the polarization angle and degree. They suggest that their results favor synchrotron radiation from a relativistic outflow with a magnetic field which is coherent on an angular size comparable with the angular size of the emitting region  (~1/$\Gamma$, where $\Gamma$ is the bulk Lorentz factor of the jet).

For now, it is safe to say that the question of the polarization of the prompt GRB emission remains open. There are hints of a large polarization, but the number of GRBs and the poor understanding of systematics makes it difficult to draw any firm conclusion. Given the importance of the subject, it is clear that the field now requires a mission dedicated to studying the polarization of the prompt GRB emission at keV-MeV energies.

At low energies, i.e. optical to radio, there have been claims of significant polarization in the early or late afterglow \cite{SM2009}\cite{RW2003}. The degree of polarization is usually low and variable. Unfortunately the data for optical polarization is scarce for the afterglow, and null for the prompt part, though an interesting attempt is made by the MASTER collaboration\cite{TL2010}. A clear, time resolved, measurement of the optical polarization would have a strong impact on the physics of the jet and internal shocks, magnetic field, and could also be used to probe fundamental physics theories such as quantum gravity. This is difficult to do however, as it requires a large enough ($\geq 1$m) telescope with appropriate instrumentation and fast reaction. At radio wavelengths, the next generation of radio telescopes such as SKA will be able to detect the polarization of both the prompt and afterglow emission of GRBs.

\subsection{The prompt emission in the optical}
\label{ss-optical}

The study of the prompt GRB emission outside the hard X-ray domain is a field which was opened in 1999, with the discovery by Akerlof et al. (1999) of the prompt optical flash produced by GRB~990123 \cite{Akerlof1999}. Such studies require the capability of measuring the position of the GRB, of distributing the alert to ground observatories, and of starting optical observations {\it before the end of the prompt phase}, that is in several seconds (Fig. \ref{fig-t90}). This became possible in the 90's thanks to the implementation of a system of alert distribution, first called BACODINE, then GCN after may 1997, at the GSFC\footnote{Goddard Space Flight Center -- USA}\cite{Barthelmy1994} and to the development of rapidly moving robotic telescopes \cite{Barthelmy1993} \cite{Boer1998} \cite{Lee1998} and wide-field optical cameras \cite{Vanderspek1992}. Starting in 2000, it has been a common feature of GRB satellites (HETE-2, Swift) and of general purpose gamma-ray missions (INTEGRAL, Agile, Fermi) to provide quick GRB localizations with an accuracy of few to several arcminutes, allowing occasional observations of the prompt emission in the optical range. These observations have revealed the diversity of the prompt optical emission of long GRBs, with some bursts being very bright at optical wavelengths and other much fainter.

\begin{figure}[htbp]
\begin{center}
\includegraphics[width=7cm]{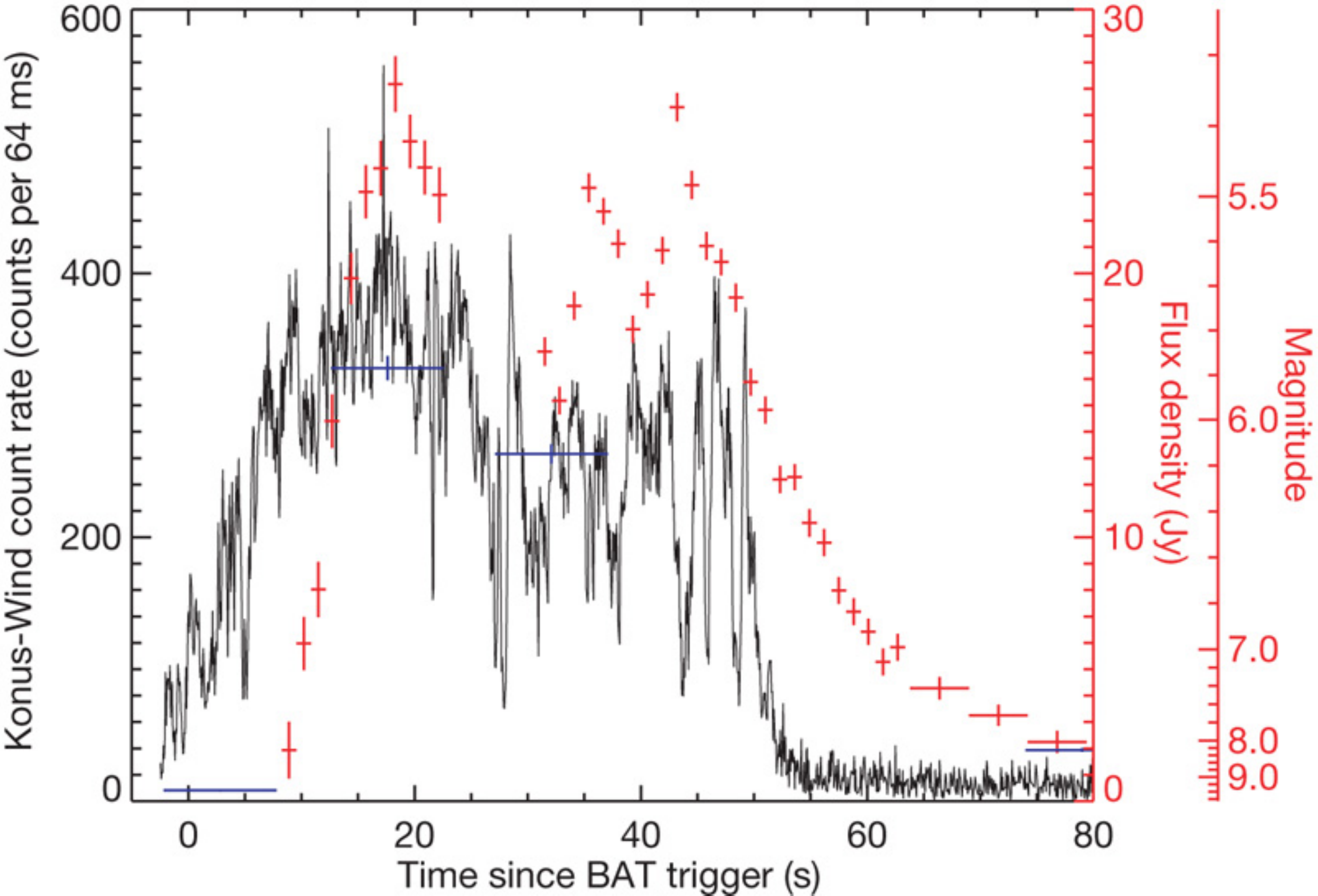}
\hskip 1.5cm
\includegraphics[width=7cm]{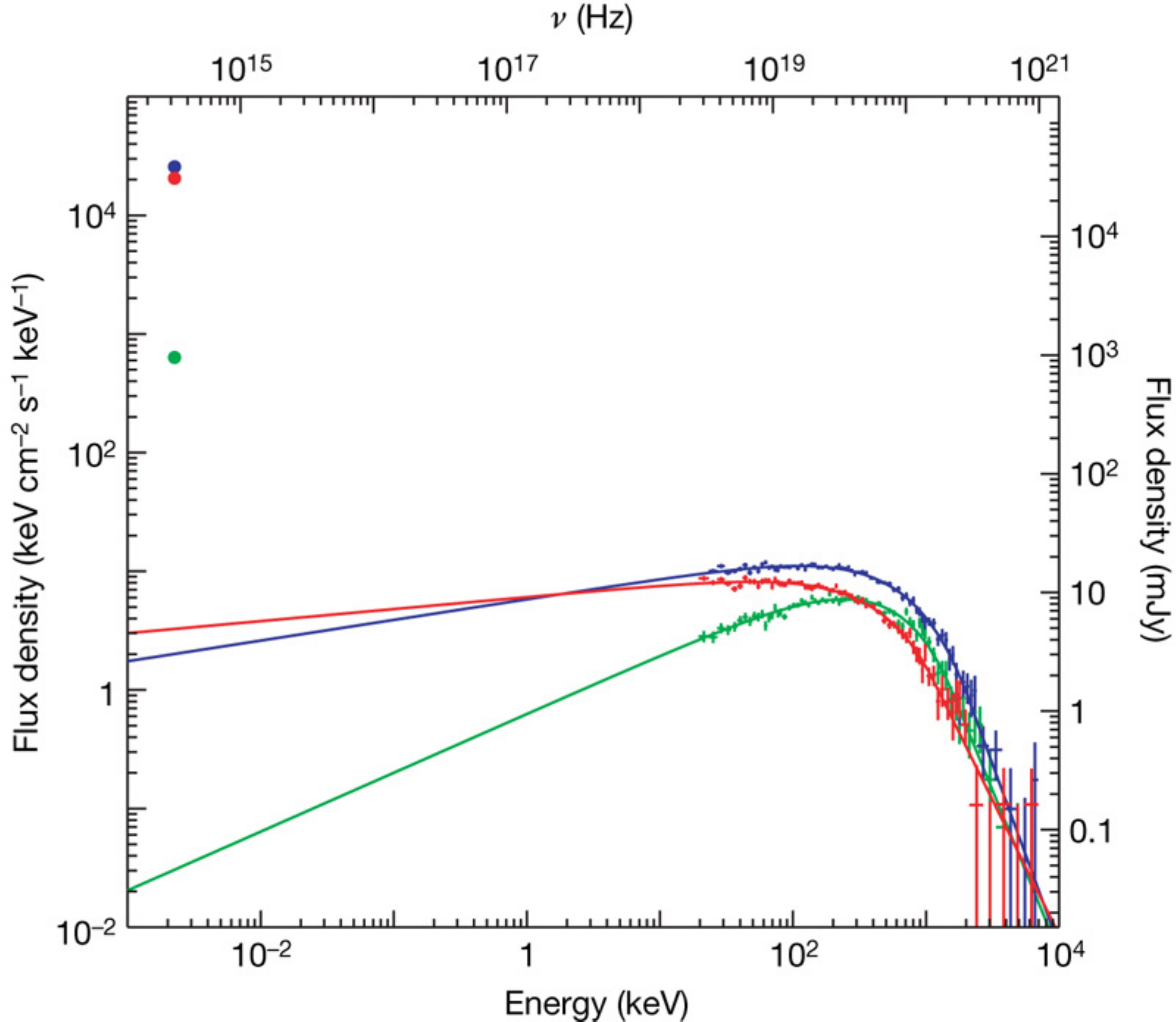}
\caption{Optical and gamma-ray light-curves (left panel) and spectral energy distribution (SED, right panel) of the very bright GRB~080319B. The optical light-curve is plotted with the red crosses and the gamma-ray flux with the black line. The right panel shows the SED measured in three 10-s time intervals during the prompt GRB, centred at T0+3 s (green), T0+17 s (blue) and T0+32 s (red). The optical emission is greatly above the extrapolation of the gamma-ray spectrum -- From Racusin et al. (2008).}
\label{fig-nakedeye}
\end{center}
\end{figure}

GRB~080319B was an optically bright burst whose prompt optical emission was observed by two wide-field instruments: 'Pi of the sky' and Tortora. These instruments were observing GRB~080319A when GRB~080319B occurred and, by chance, the two GRB were sufficiently close in time (separated by 30 minutes) and on the sky (separated by 10$^\circ$), to allow them to observe GRB~080319B at optical wavelength, even before the reception of Swift alert. The light-curve presented in figure \ref{fig-nakedeye} shows a bright optical flare (m$_{\rm V} \sim 5.4$ at maximum) lasting about 50~sec, with a duration very similar to the prompt emission in gamma-rays. This optical flare is quickly variable and it rises $\sim$10 seconds after the beginning of the gamma-ray burst. The spectral energy distribution in the right panel of figure \ref{fig-nakedeye} shows that the optical emission is well above the extrapolation of the gamma-ray spectrum, implying that it comes from a different emission mechanism or from a different region of emission. This behavior seems quite common in GRBs.

\begin{figure}[htbp]
\begin{center}
\includegraphics[width=7cm]{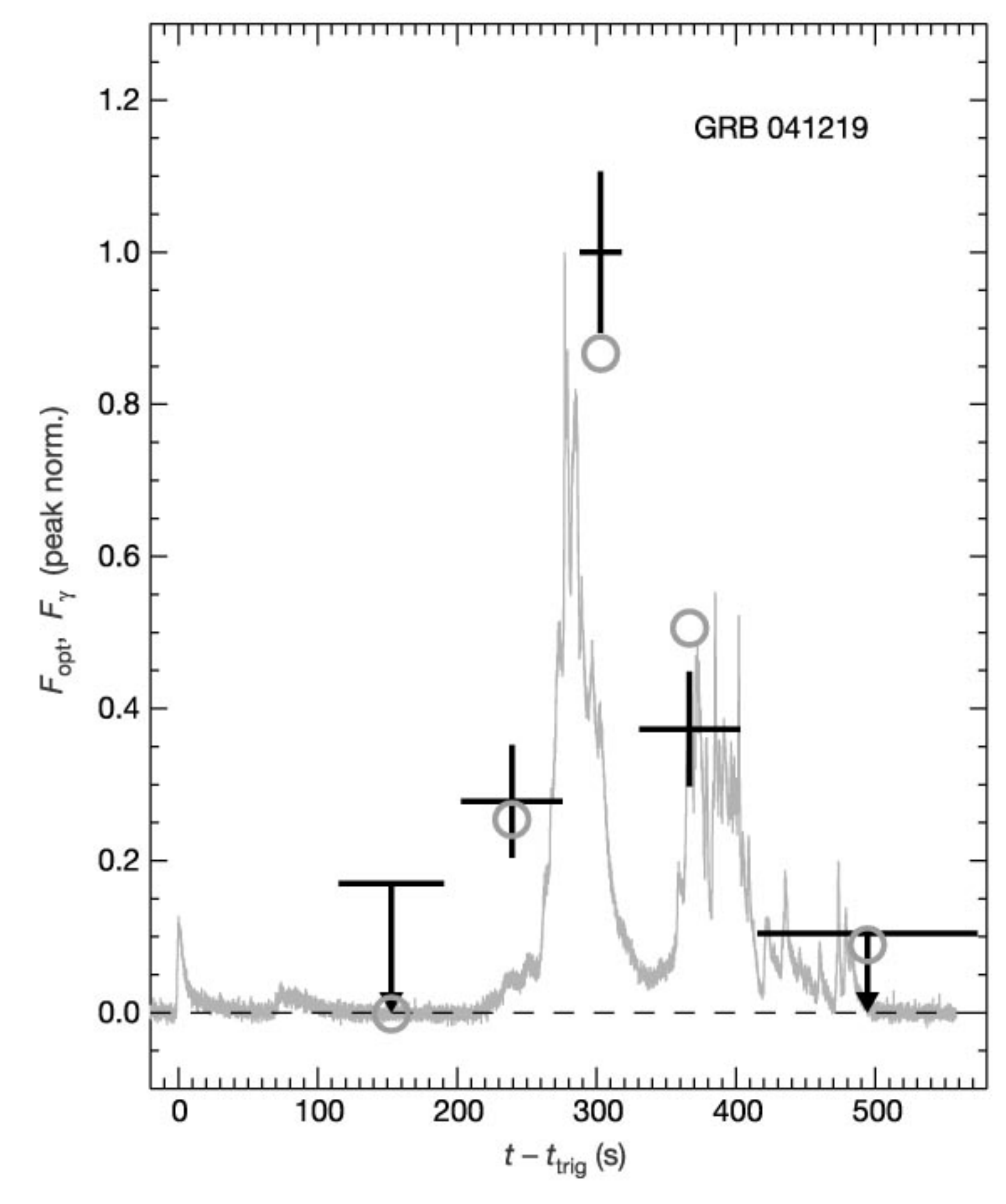}
\hskip 1.5cm
\includegraphics[width=7cm]{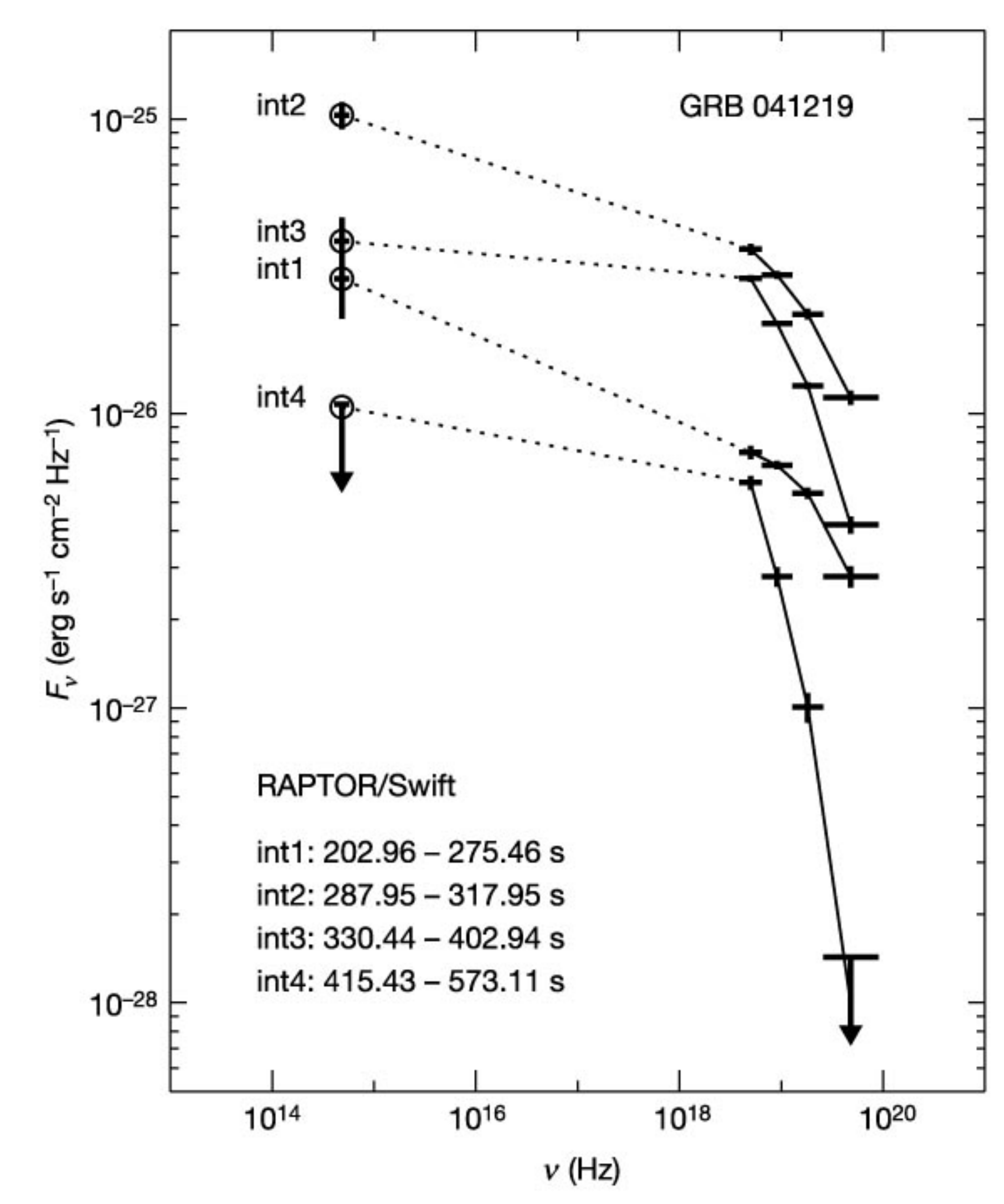}
\caption{Optical and gamma-ray light-curves (left panel) and spectral energy distribution (SED, right panel) of the very bright GRB~041219A. The black crosses show the optical flux measured by RAPTOR \cite{Vestrand2005}, and the continuous grey line is the gamma-ray light-curve. The circles indicate the optical flux that would be expected for a constant optical/gamma-ray flux ratio. The right panel displays the optical to gamma-ray SED measured in 4 time intervals. This panel shows that the optical and gamma-ray flux evolve in concert during the prompt phase of the burst -- From Vestrand et al. (2005).}
\label{fig-vis-gamma}
\end{center}
\end{figure}

GRB~041219A shows a completely different case. The main episode of gamma-ray emission of this burst was observed by various optical and NIR instruments thanks to a precursor emitted about 250 seconds earlier \cite{Vestrand2005}\cite{Blake2005}\cite{McBreen2006}. Figure \ref{fig-vis-gamma} shows that the optical and gamma-ray fluxes evolve in concert during the prompt phase (right panel) and that the optical emission is more or less a constant fraction of the gamma-rays. In the case of GRB~041219A, it is thus reasonable to assume that the optical emission is simply the low energy end of the gamma-ray spectrum.

Yost et al. (2007) undertook a systematic analysis of the few GRBs with prompt optical observations in order to determine the fraction exhibiting extra optical emission (beyond the extrapolation of the prompt gamma-ray spectrum). They compared the (low energy) slope of the gamma-ray spectrum, $\beta_{\gamma}$, with the slope of a line connecting the optical and gamma-ray flux measurements, $\beta_{(o-\gamma)}$. GRBs with $\beta_{(o-\gamma)} < \beta_{\gamma}$ indicate an excess of optical emission, while $\beta_{(o-\gamma)} > \beta_{\gamma}$ indicate a deficit of optical photons with respect to the simple extrapolation of the gamma-ray spectrum. This second case can be easily explained with a spectral break between the gamma-ray and the optical frequencies or with some extinction of the GRB optical light.\footnote{Optical and gamma-ray photons undergo very different absorption along the line of sight. While gamma-ray photons beyond few keV travel almost freely (until they encounter a man-made detector...), this is not the case of optical photons which can be strongly absorbed by dust both in the host galaxy and in our galaxy.} Among 7 GRBs, they find 2 bursts showing evidence of prompt optical emission in excess of the extrapolation of the gamma-ray spectrum: GRB~990123 and GRB~050904 (see also GRB~080319B discussed above), 2 bursts where the optical emission is compatible with the extrapolation of the gamma-ray spectrum, and 3 bursts where a spectral break or optical extinction in the host galaxy could explain a prompt optical flux lower than the extrapolation of the gamma-ray spectrum. GRB~041219a, for instance, displays nice evidence for an evolving spectral break between optical and the gamma-ray frequencies, while GRB~050401 undergoes significant optical extinction within its host galaxy \cite{Watson2006}.
To conclude, the close resemblance of the prompt emission of GRB~080319B at gamma-ray and optical frequencies, suggests that both components were emitted in a single region by different radiation processes, however this interpretation contradicts previous explanations, which tended to attribute the prompt optical emission to the reverse shock. It is probably fair to say that the small number of events for which prompt optical emission has been detected with good significance and good temporal resolution does not allow to draw firm conclusions about the origin of this emission. It is one of the challenges of GRB science in the coming years to be able to measure the prompt optical (visible and NIR) emission of a significant fraction of detected GRBs (see Section \ref{ss-svom} below).

\begin{figure}[htbp]
\begin{center}
\includegraphics[width=8cm]{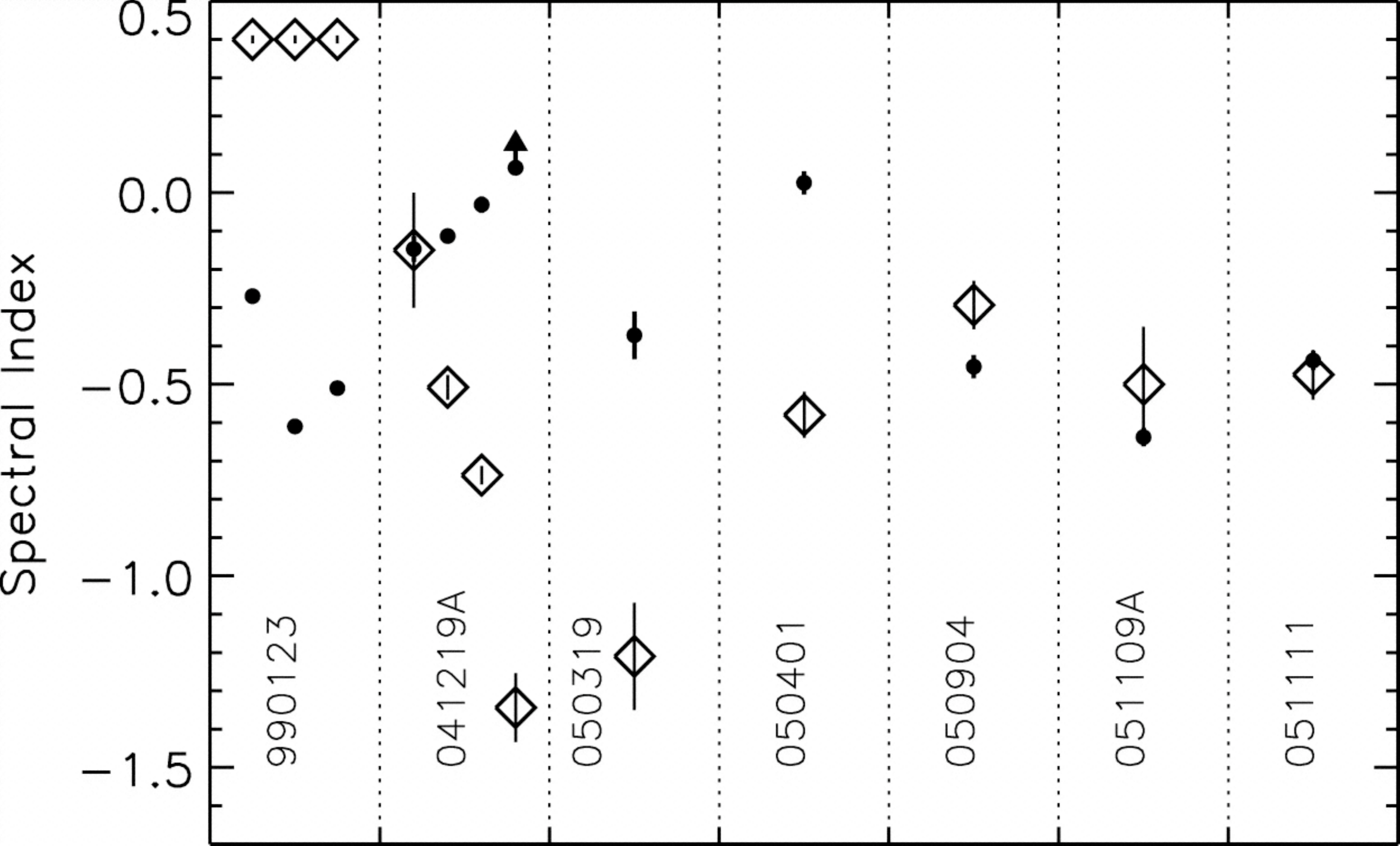}
\caption{ Comparison of prompt optical-to-gamma-ray spectral indices ($\beta_{(o-\gamma)}$, solid circles) to the spectral indices within the gamma-ray band ($\beta_{\gamma}$, open diamonds). Multiple measurements are available for GRB~990123 and GRB~041219A. GRB~990123 and GRB~050904 have $\beta_{(o-\gamma)} < \beta_{\gamma}$, indicating an excess of optical emission above the extrapolation of the gamma-ray spectrum at low energies. -- From Yost et al. (2007).}
\label{fig-rotse}
\end{center}
\end{figure}

\subsection{The prompt emission in 2010}
\label{ss-issues}

This brief overview shows that our understanding of the prompt GRB emission has made great progress in the recent years, even while the focus of GRB work was more towards the study of the afterglows, progenitors, and host galaxies. We are now aware of the broad luminosity function of GRBs, of the existence of relations linking the luminosity with observed parameters like the variability or E$_{\rm peak}$, of the complexity of the panchromatic spectrum of GRBs, and of the need to study their polarization...

Despite these advances, tantalizing questions remain about the prompt GRB phase: the physics of the prompt phase, the origin of the radiation, the emitting regions contributing to the prompt GRB, the physical conditions within the jet, the connection between the properties of the jet (or the progenitor) and the properties of the prompt GRB (duration, E$_{\rm peak}$), the shape and extent of the luminosity function...
We quickly discuss below how SVOM and future missions will help addressing some of these questions.

\section{The prompt emission in the SVOM era}
\label{s-svom}

One strength of SVOM is that it will be part of a very rich panorama of astronomical instruments. In the period 2015-2020 and beyond, powerful instruments will be available to follow-up SVOM GRBs. On the ground, the list includes the second generation instruments on the VLTs (especially the large band spectrograph X-Shooter), the Extremely Large telescopes in Europe and in the US, the precursors of the Square Kilometer Array at radio frequencies, the Cerenkov Telescope Array at TeV energies, and a new generation of gravitational waves and neutrino detectors. For the first time, the direct detection of GRBs from the ground may become possible, with large radio arrays (e.g. LOFAR) and with the discovery of young GRB afterglows (age $\le$1 day) with the Large Synoptic Survey Telescope without need of a high-energy trigger. The complementarity of SVOM with LOFAR and the LSST, will undoubtedly provide a fruitful new view on GRBs. In space, the Fermi telescope may continue to operate and the James Webb Space Telescope will give us an unprecedented view on the host galaxies of distant GRBs.

The observing strategy of SVOM has been designed to permit the systematic follow-up of Swift GRBs (see the contribution of J. Paul et al. in this volume) with large telescopes on Earth. Considering this strategy, the capability of SVOM for rapid afterglow localization at visible and NIR wavelengths with the telescopes of SVOM,\footnote{The Visible Telescope on-board the spacecraft and the two Ground Follow-up Telescopes on Earth \cite{Paul2010}.} the availability of new powerful spectrographs, and the continuing interest for GRBs in the next decade, we rely on the fact that 50-60\% of SVOM GRBs will have their redshift measured (compared with 30\% for Swift). This is very important since redshifts are needed to measure several fundamental characteristics of GRBs required to understand these events, like their overall energetics or their intrinsic properties. 

GRBs with no redshift, on the other hand, are useful for general statistical studies. Even in the absence of a redshift, the detection of an infrared prompt counterpart can be the signature of an evolving surrounding dust medium, and polarization can give meaningful physical parameters. The detection of radio counterparts is also interesting, since it may lead to the identification of the host, and in some cases allow GRB calorimetry \cite{Frail2005}.

\subsection{Studying the prompt GRB phase with SVOM}
\label{ss-svom}

The science case of SVOM emphasizes the study of the physics of GRBs \cite{Paul2010}. This has led the SVOM team to include in the mission three instruments designed to catch and observe the prompt emission with good sensitivity and over a broad range of frequencies. ECLAIRs and the Gamma-Ray Monitor will jointly cover energies from 4 keV to 5 MeV, with an overlap in the range 30-150 keV. This combination will permit the measure of GRB spectra and light-curves over a broad energy range, giving access to some fundamental properties of GRBs like their duration, fluence, \epeak... For about 20\% of the GRBs, the GWAC will provide simultaneous observations at optical frequencies, allowing much more systematic studies of the panchromatic spectrum of the prompt emission and a better understanding of the emitting regions and of the radiation processes at work in GRBs. The continuous energy coverage between ECLAIRs and the follow-up X-ray telescope (MXT, the Micro-channel X-ray Telescope), will permit detailed observations of the transition period between the prompt GRB and the early afterglow, when the radiation from different regions overlap. For about half the bursts, the GFTs will also observe this crucial period at visible and near infrared frequencies. These powerful diagnostics may be complemented by one of the dedicated GRB polarimetry missions that are currently under study (e.g. POLAR, POET...).

The good sensitivity of SVOM at low energies (with an effective area $\ge$150 cm$^2$ at 4 keV) will permit scratching a little more the population of soft sub-luminous GRBs. SVOM will thus provide new constraints on the GRB luminosity function and it will open a new window of a possible population of nearby faint and soft GRBs.

\subsection{Detecting the prompt GRB emission without the help of a spacecraft}
\label{ss-signals}

We are entering a remarkable epoch for GRB studies, with the emerging opportunity to detect GRBs directly from the ground without the help of a spacecraft. While the LSST will detect the optical afterglows of GRBs during the first day(s) after the burst, providing no information about the prompt emission, other projects, like LOFAR in radio will have the capability to detect short radio transients due to GRBs, if they exist. In a completely different domain, Advanced LIGO and Advanced VIRGO will have a good chance to detect the gravitational chirps produced by the coalescence of two compact stars, which may also produce short GRBs. IceCUBE or KM3NeT will search for transient signals possibly associated with GRBs, opening the multi-messenger window. 
SVOM will play a crucial role in this new context, as it will contribute to identify which of these new signals are truly associated with GRBs.

\subsection{Other challenges}
\label{s-challenges}

Despite the exceptional coverage of the prompt emission and the progress awaited with SVOM (section \ref{ss-svom}), some observations will remain out of reach. The most exciting concern probably the measure of the polarization of the prompt GRB at X-ray, gamma-ray and optical frequencies, and a deep exploration of the realm of faint transients at keV energies. These objectives require dedicated instrumentation which has not flown yet, but which is under development, like wide-field hard X-ray and gamma-ray polarimeters (e.g. POLAR \cite{Lamanna2008}, POET \cite{McConnell2009}, GRIPS \cite{Greiner2009}, GAP \cite{Yonetoku2010}) or focusing X-ray optics (e.g. lobster eye optics \cite{Hudec2010}). These new devices will complement SVOM panchromatic view of the prompt emission, promising a new understanding of this enigmatic phase of GRBs during which they radiate most of their power.




\end{document}